\documentclass[aps,prd,showpacs,twocolumn,nofootinbib,superscriptaddress,amsmath,amssymb]{revtex4}

\usepackage{graphics}
\usepackage{epsfig}
\usepackage{graphicx}
\usepackage{color}
\usepackage{mathrsfs}
\usepackage{bm}

\usepackage{amsmath}

\newcommand{\rev}[1]{{#1}}
\newcommand{\rv}[1]{{#1}}

\begin{document}
\title{A Note on Bloch theorem}
\author{C.X. Zhang}
\affiliation{Physics Department, Ariel University, Ariel 40700, Israel}

\author{M.A. Zubkov \footnote{On leave of absence from Institute for Theoretical and Experimental Physics, B. Cheremushkinskaya 25, Moscow, 117259, Russia}}
\email{zubkov@itep.ru}
\affiliation{Physics Department, Ariel University, Ariel 40700, Israel}

\date{\today}

\begin{abstract}
Bloch theorem in ordinary quantum mechanics means the absence of the total electric current in equilibrium. In the present paper we analyze the possibility that this theorem remains valid within quantum field theory relevant for the description of both  high energy physics and condensed matter physics phenomena. First of all, we prove that the total electric current in equilibrium is the topological invariant for the gapped fermions that are subject to periodical boundary conditions, i.e. it is robust to the smooth modification of such systems. This property remains valid when the inter - fermion interactions due to the exchange by bosonic excitations are taken into account perturbatively. We give the proof of this statement to all orders in perturbation theory. Thus we prove the weak version of the Bloch theorem, and conclude that the total current remains zero in any system, which is obtained by smooth modification of the one with the gapped charged fermions, periodical boundary conditions, and vanishing total electric current. We analyse several examples, in which the fermions are gapless. In some of them the total electric current vanishes. At the same time we propose the counterexamples of the equilibrium gapless systems, in which the total electric current is nonzero.
\end{abstract}


\maketitle

\section{Introduction}

According to the conventional quantum mechanical formulation of the Bloch theorem \cite{B1_} in the infinitely large equilibrium system the total electric current is zero. The proof of this theorem is known within the framework of ordinary quantum mechanics with fixed finite number of particles. There were several attempts to generalize  the Bloch theorem to quantum field theory (QFT)\footnote{Quantum field theory represents both the mathematical basis for the description of the high energy physics and the condensed matter physics. The difference between the corresponding models is actually in symmetry. The high energy physics systems obey Lorentz invariance while typical condensed matter systems do not. Aside from the Lorentz symmetry the description of the condensed matter systems and the description of the high energy physics systems are completely equivalent. If we consider lattice regularization of the high energy physics models, then the analogy becomes more complete. }. However, this extension has been limited so far by the consideration of specific models. For example, a continuum model in the presence of magnetic field has been considered in \cite{B2_} while some lattice models were discussed in \cite{B4_,B5_,B6_}. In \cite{B3_} the attempt was made to prove the Bloch theorem for the QFT of general type. The proof presented in \cite{B3_}, however, seem to us not clear enough. Moreover, below we present the example of the QFT system, in which the Bloch theorem in its conventional formulation does not work. In the recent paper \cite{Watanabe} the proof of the Bloch theorem has been presented for the arbitrary lattice one - dimensional model. This proof may also be extended to the higher dimensional lattice systems, which are infinite in one particular direction, and are compact in the other directions. In this setup Bloch theorem states the absence of persistent current in the direction, in which the system is infinite. The same form of the Bloch theorem has been proposed in \cite{VolovikBloch}. The possible extension of the Bloch theorem to the QFT may be important for the applications of the QFT techniques to the condensed matter physics (see, e.g. \cite{V1,V2,V3,V4,V5,V6,V7,S1,S2,S3,S4,S5,S6,S7,B1,B2,B3,B4,B5,B6,G1,G2,G3}).

In the present paper we analyse the possible form in which Bloch theorem survives in  an infinite fermionic QFT system. First of all, we demonstrate that in the conventional formulation the Bloch theorem does not hold: we present the example of an infinite system, in which there is the persistent current in equilibrium. Instead of the conventional Bloch theorem we prove its weakened version. It states, that the total electric current in the equilibrium infinite system with periodical spatial boundary conditions and gapped charged fermions is the topological invariant, i.e. it is not changed when the system is modified smoothly. The whole set of the gapped QFT systems may be divided into the homotopic classes. Within each class the systems are connected by continuous modification. Therefore, if the total electric current vanishes in one of such systems, it vanishes in all systems that belong to the same homotopic class.
On the technical side we will use Wigner - Weyl formalism \cite{1,2,berezin,6} adapted in \cite{Z2016,Z2016_1,FZ2019,SZ2018,ZK2017,KZ2018,KZ2017,ZK2018,ZZ2019_2,ZKA2018,AKZ2018} to the lattice models of solid state physics combined with the ordinary perturbation theory.
In the presence of the external fields, which break the translational symmetry \cite{Zhang+Zubkov_2019_Feynman}, the Wigner transformation of Green functions is more useful for our purposes than the Fourier transformation. Using the technique of Wigner transformation we express the response of electric current to the external electromagnetic field. The Feynman diagrams written in terms of the Wigner - transformed Green functions contain the same amount of integrations over momenta as the Feynman diagrams in the homogeneous theory. This facilitates considerably the calculations.
At the moment  we cannot establish any definite analogue of the Bloch theorem
for the gapless QFT systems of general type.
Instead we analyse several particular examples, where Bloch theorem holds/does not hold.

The paper is organized as follows. In Sect. \ref{Sect2} we describe the formulation of  fermionic QFT models using Wigner - Weyl formalism. In Sect. \ref{Sect4} we present the proof that in the noninteracting gapped fermionic systems with periodical spacial boundary conditions the total electric current is the topological invariant. In Sect. \ref{Sectex0} we demonstrate that the conventional lattice models with noninteracting gapped fermions have vanishing total current. Notice, that the perturbative  inclusion of interactions via the exchange by neutral bosons does not change this conclusion. The proof is given in Sect. \ref{SectCoulomb}. In Sections \ref{SectE1}, \ref{SectE2} and \ref{SectE3} we consider the particular gapless systems. In Sect. \ref{SectE1} we discuss the typical example of the system, in which the Bloch theorem holds. In Sect. \ref{SectE2} we consider the counter - example of the equilibrium system, in which the Bloch theorem does not hold. Notice, that this system does not satisfy the additional conditions needed for the validity of Bloch theorem proposed in \cite{B3,VolovikBloch}. In Sect. \ref{SectE3} we discuss the example of the system that obeys Bloch theorem at a certain range of the values of Fermi energy and does not obey it for another ranges of the Fermi energy. In Sect. \ref{SectConcl} we end with conclusions.

\section{Gapped fermions}

\subsection{Noninteracting fermions and Wigner - Weyl formalism}

\label{Sect2}

Let us consider the continuum system of non - interacting particles. In the homogeneous case the partition function of such a system in momentum space has the form \cite{Z2016,Z2016_1,SZ2018}
{\begin{eqnarray}
Z &=& \int D\bar{\psi}D\psi \, {\rm exp}\Big(  \int_{\cal M} \frac{d^D {p}}{|{\cal M}|}\nonumber\\&&\bar{\psi}^T({p}){Q}(p)\psi({p}) \Big),\label{Z01}
\end{eqnarray}}
Here $|{\cal M}| = (2\pi)^D$, where $D$ is the dimension of space - time. $\bar{\psi}$ and $\psi$ are the anticommuting multi - component Grassmann variables defined in momentum space. Matrix  $Q(p)$ is given by
$$
Q = i \omega - \hat{H}(\bf p)
$$
Here $p = (\omega, {\bf p})$.  Introduction of the external gauge field $A(x)$ defined as a function of coordinates results in Peierls substitution (see, for example, \cite{Z2016,Z2016_1,SZ2018}):
{\begin{eqnarray}
Z &=& \int D\bar{\psi}D\psi \, {\rm exp}\Big(  \int_{\cal M} \frac{d^3 {p}}{|{\cal M}|} \nonumber\\&&\bar{\psi}^T({p}){Q}(p - A(i\partial_p))\psi({p}) \Big),\label{Z01}
\end{eqnarray}}
where the products of operators inside expression ${\cal Q}(p - A(i\partial_p))$ are symmetrized.

We denote for the operators $\hat{Q} = Q(p-A(i\partial_p))$ and $\hat{G} = \hat{Q}^{-1}$ their matrix elements by ${\cal Q}(p,q)$ and ${\cal G}(p,q)$ correspondingly:
$$
{\cal Q}(p,q) = \langle p|\hat{Q}| q\rangle, \quad {\cal G}(p,q) = \langle p|\hat{Q}^{-1}| q\rangle
$$
where $|p\rangle$ and $|q\rangle$ are momentum eigenstates.
The basis of Hilbert space of functions is normalized as $\langle p| q\rangle = \delta^{(D)}(p-q)$. The mentioned operators satisfy
$$
\langle p|\hat{Q}\hat{G}|q\rangle = \delta^{(D)}({ p} - { q}).
$$
We insert here the complete set of momentum eigenstates \rv{ $\{ |k\rangle \}$, and obtain
\begin{eqnarray}\label{QG_momentum}
\int {\cal Q}(p,k) {\cal G}(k,q) dk=\delta(p-q).
\end{eqnarray}  }
Eq. (\ref{Z01}) may be rewritten as follows
{\begin{eqnarray}
Z &=& \int D\bar{\psi}D\psi \, {\rm exp}\Big(  \int_{\cal M} \frac{d^3 {p}_1}{\sqrt{|{\cal M}|}} \int_{\cal M} \frac{d^3 {p}_2}{\sqrt{|{\cal M}|}}\nonumber\\&&\bar{\psi}^T({p}_1){\cal Q}(p_1,p_2)\psi({p}_2) \Big),\label{Z1}
\end{eqnarray}}
while the Green function is
{\begin{eqnarray}
{\cal G}_{ab}(k_2,k_1)&=& \frac{1}{Z}\int D\bar{\psi}D\psi \, {\rm exp}\Big(  \int_{\cal M} \frac{d^3 {p}_1}{\sqrt{|{\cal M}|}} \int_{\cal M} \frac{d^3 {p}_2}{\sqrt{|{\cal M}|}}\nonumber\\&&\bar{\psi}^T({p}_1){\cal Q}(p_1,p_2)\psi({p}_2) \Big) \frac{\bar{\psi}_b(k_2)}{\sqrt{|{\cal M}|}} \frac{\psi_a(k_1)}{\sqrt{|{\cal M}|}}\label{G1}
\end{eqnarray}}
Indices $a,b$ enumerate the components of the fermionic fields, which will be omitted later for simplicity.

The Wigner transformation of $\cal G$ is defined as:
\begin{equation}
\begin{aligned}
{G}_W(x,p) \equiv \int dq e^{ix q} {\cal G}({p+q/2}, {p-q/2})\label{GWx}
\end{aligned}
\end{equation}
Here integral is over $q$ that belong to momentum space. Weyl symbol of operator $\hat{Q}$ is defined in the similar way:
$$
{Q}_W(x,p) \equiv \int dq e^{ix q} {\cal Q}({p+q/2}, {p-q/2})
$$
In the following we will use the following identity of Wigner - Weyl formalism: if $C(p_1,p_2)=\int A(p_1,q)B(q,p_2)dq$ then the Wigner transformations of $A,B,C$ obey
 $C_W(x,p)= A_W(x,p)\star B_W(x,p)$.
In continuous theory, from Eq.(\ref{QG_momentum}) if follows that
the Weyl symbol of $\hat{Q}$ and the Wigner transformation
of $\cal G$ obey the Groenewold equation (see, for example, \cite{Z2016,Z2016_1,SZ2018})
\begin{equation}\begin{aligned}
&1 = {G}_W(x,p) \star Q_W(x,p) \\&=
{G}_W(x,p)
e^{\frac{i}{2} \left( \overleftarrow{\partial}_{x}\overrightarrow{\partial_p}-\overleftarrow{\partial_p}\overrightarrow{\partial}_{x}\right )}
Q_W(x,p)
\label{GQW}\end{aligned}\end{equation}
This equation is also satisfied for the case of the lattice model with compact Brillouin zone provided that the fields entering $\hat{Q}$ vary slowly, i.e. their variations at the distance of the order of the lattice spacing may be neglected \cite{FZ2019,ZZ2019_2}. In practise this requirement is always satisfied in the real solids if the inhomogeneity is caused by external magnetic field or by elastic deformations.
In spite of the appearance of the complicated star products,
the use of Wigner-transformed Green function $G_W$ has certain advantages
compared to the use of the ordinary momentum space Green function $G(p,q)$.
Feynman diagrams written in terms of $G_W$ are more concise. In what follows we will see, that these expressions contain the same amount of integrations over momenta as the Feynman diagrams of the homogenous theory.

The Grassmann - valued Wigner function may be defined as
{$$
W(p,q) = \frac{\bar{\psi}(p)}{\sqrt{|{\cal M}|}}\frac{{\psi}(q)}{\sqrt{|{\cal M}|}}
$$}
We may define operator $\hat{W}[\psi,\bar{\psi}]$, whose matrix elements are equal to $W(p,q) = \langle p|\hat{W}[\psi,\bar{\psi}]|q\rangle$.

If the field $A$ is slowly varying then \rv{ ${Q}_W(p,x)= Q_W(p-{ A}(x))$ \cite{SZ2018}.}
As a result the partition function receives the form:
{\begin{eqnarray}
Z& =&
\int D\bar{\psi}D\psi \, e^{  \sum_x \int \frac{dp}{(2\pi)^D} {\rm Tr} {W}_W[\psi,\bar{\psi}](p,x) \star Q_W(p,x)}\nonumber
\end{eqnarray}}
where by $W_W$ we denote the Weyl symbol of $\hat{W}$.

\subsection{Equilibrium current as topological invariant for the gapped noninteracting  fermions}

\label{Sect4}

In this section we consider gapped noninteracting charged fermions in the presence of periodical boundary conditions.
Let us consider the variation of the partition function
{\begin{eqnarray}
&&\delta {\rm log} \,Z =- \frac{1}{Z}
\int D\bar{\psi}D\psi \, {\rm exp}\Big(  \sum_x \int \frac{dp}{(2\pi)^D}\nonumber\\ && {\rm Tr} {W}_W[\psi,\bar{\psi}](p,x) \star Q_W(p - { A}(x)) \Big) \nonumber\\
&& \sum_x \int \frac{dp}{(2\pi)^D} {\rm Tr} {W}_W[\psi,\bar{\psi}](p,x)\nonumber\\ && \star \partial_{p_k} Q_W(p - { A}(x))\delta{ A}_k(x)  \nonumber\\
&& \approx - \frac{1}{Z}
\int D\bar{\psi}D\psi \, {\rm exp}\Big(  \int_x \int \frac{dp}{(2\pi)^D} {\rm Tr} {W}_W[\psi,\bar{\psi}](p,x) \nonumber\\
&& \star Q_W(p - { A}(x)) \Big) \nonumber\\
&& \int dx \int \frac{dp}{(2\pi)^D} {\rm Tr} {W}_W[\psi,\bar{\psi}](p,x)\nonumber\\
&& \star \partial_{p_k} Q_W(p - { A}(x))\delta{ A}_k(x) \label{dZ2}
\end{eqnarray}}
The current density integrated over the whole volume of the system appears as the response to the variation of $A$:
\begin{eqnarray}
&&\langle J^k \rangle =  -\frac{T}{Z}
\int D\bar{\psi}D\psi \, {\rm exp}\Big(  \int_x \int \frac{dp}{(2\pi)^D} {\rm Tr} {W}_W[\psi,\bar{\psi}](p,x) \nonumber\\
&& \star Q_W(p - { A}(x)) \Big) \nonumber\\
&& \int d^D x \, \int \frac{dp}{(2\pi)^D}\, {\rm Tr} {W}_W[\psi,\bar{\psi}](p,x)  \partial_{p_k} Q_W(p - { A}(x)) \nonumber\\
&&= -T\, \int d^D x \, \int \frac{dp}{(2\pi)^D}\, {\rm Tr} \, G_W(p,x)  \partial_{p_k} Q_W(p - { A}(x))\label{J}
\end{eqnarray}
In the presence of periodic boundary conditions the properties of the star product allow to rewrite the last equation in the following way:
{\begin{eqnarray}
\langle J^k \rangle
&=& - T\, \int d^D x \, \int \frac{d^Dp}{(2\pi)^D}\,\nonumber\\
&& {\rm Tr} \, G_W(p,x) \star \partial_{p_k} Q_W(p -{ A}(x))  \label{J2}
\end{eqnarray}}
Provided that there are no divergencies in this expression, it is the topological invariant, i.e. it is not changed when the system is modified continuously.  The proof is as follows. \rv{Let us consider an arbitrary variation
$Q_W \rightarrow Q_W + \delta Q_W$, and the related variation
of the Green function:
$G_W \rightarrow G_W + \delta G_W$, according to Eq.(\ref{GQW}).}
The variation of electric current receives the form
\begin{eqnarray}
&&\delta J^{k}
=- T\,\delta \int_{} \, d^D x \, \int \frac{d^D p}{(2\pi)^D}\, {\rm Tr} G_W\star  \partial_{p_k}  Q_W \nonumber\\
&&= - T\,\int_{} \,d^D x \, \int \frac{d^D p}{(2\pi)^D}\, {\rm Tr} (\delta G_W \star  \partial_{p_k}  Q_W+G_W\star  \partial_{p_k}  \delta Q_W)
\nonumber\\
&&=- T\, \int_{} \, d^D x \, \int \frac{d^D p}{(2\pi)^D}\,{\rm Tr} (-G_W\star  \delta Q_W \star  G_W  \star  \partial_{p_k}  Q_W\nonumber\\&&+G_W\star  \partial_{p_k}  Q_W)\nonumber\\
&&=- T\, \int_{} \, d^D x \, \int \frac{d^D p}{(2\pi)^D}\,{\rm Tr} (\delta Q_W \star  \partial_{p_k}  G_W +G_W \star  \partial_{p_k}  \delta Q_W)\nonumber\\
&&= - T\,\int_{}d^D x \, \int \frac{d^D p}{(2\pi)^D} \, \partial_{p_k} \, {\rm Tr} (\delta Q_W \star   G_W)=0\,,
\end{eqnarray}
In the last step we assume the periodical boundary conditions in $p$ space. This occurs, in particular, for the lattice tight - binding model with compact Brillouin zone.
In the above proof it has also been implied, that the integrals are convergent. This assumes both the absence of ultraviolet and infrared divergencies. The latter are absent if neither $\hat{G}$ nor $\hat{Q}$ have poles. The ultraviolet divergencies may be eliminated if the theory on the lattice (the tight - binding model) is considered. This guarantees that the ultraviolet divergencies at large spacial momenta are absent. For the noninteracting system with ${\hat Q} = i\omega  - \hat{H}$ the uncertainty in the integral over $\omega$ remains. However, if the aim is to calculate the conductivity (i.e. the response of Eq. (\ref{J2}) to the external electric field), then the corresponding integral in $\omega$ that follows from Eq. (\ref{J2}) is convergent at $\omega \to \infty$ because the expression in the integral behaves as $\frac{1}{\omega^s}$ with $s > 1$ at $\omega \to \infty$. The integral over $\omega$ is to be regularized if we are interested in the expression for the current out of the linear response to external gauge field. The standard regularization used for the calculation of various vacuum averages of the bilinear combination of operators results in the modification $G_W(p,x) \to G_W^{(reg)}(p,x) = e^{i\tau \omega} G_W(p,x)$, where $\tau$ is to be set to $\tau \to +0$ at the end of calculations. Correspondingly, we regularize $Q_W(p,x) \to Q_W^{(reg)}(p,x) = e^{-i\tau \omega} Q_W(p,x)$. This regularization allows to save the topological invariance of the regularized Eq. (\ref{J2}). The integral $\int d\omega \frac{e^{i\omega \tau}}{i\omega - {\cal E}_n}$ entering Eq. (\ref{J2}) (here ${\cal E}_n$ is the $n$ - th eigenvalue of the Hamiltonian) may be calculated using the residue theorem:
$$
\int d\omega \frac{e^{i\omega \tau}}{i\omega - {\cal E}_n}\Big|_{\tau = +0} = 2\pi \theta(-{\cal E}_n)
$$
provided that ${\cal E}_n \ne 0$. In the case when  the value of ${\cal E}_n$ vanishes this integral remains infrared divergent.

We come to the conclusion that the total electric current is not changed if the gapped system with compact Brillouin zone is modified smoothly. This conclusion holds also when the interactions between the fermions are taken into account (see below Sect. \ref{SectCoulomb}).

\subsection{An example of the system with vanishing total current}

\label{Sectex0}
Let us discuss the case of the noninteracting fermions with Hamiltonian $\hat{H}$. Then
$$
\hat{Q} = i\omega - \hat{H}
$$
For the case of the homogeneous system with $\hat{H} = H(\hat{p})$, Eq. (\ref{J2}) receives the form
\begin{eqnarray}
\langle J^k \rangle &=&  V \, \int \frac{d\omega d^{D-1} p}{(2\pi)^D}\, {\rm Tr} \, \frac{1}{i\omega - H(p)}  \partial_{p_k} H(p)  \label{J2_0}
\end{eqnarray}
The integral over $\omega$ is to be regularized: $G_W(p,x) \to G_W^{(reg)}(p,x) = e^{i\tau \omega} G_W(p,x)$, where $\tau \to +0$. The integral entering Eq. (\ref{J2_0}) may be calculated using the residue theorem:
$$
\int d\omega \frac{e^{i\omega \tau}}{i\omega - {\cal E}_n}\Big|_{\tau = +0} = 2\pi \theta(-{\cal E}_n)
$$
provided that ${\cal E}_n \ne 0$. We have
\begin{eqnarray}
\langle J^k \rangle &=&  V \, \int \frac{ d^{D-1} p}{(2\pi)^{D-1}}\, {\rm Tr} \,  \theta(- H(p))  \partial_{p_k} H(p)  \label{J2__0}
\end{eqnarray}
For the case of the periodical boundary conditions in momentum space (say, for the lattice model with compact Brillouin zone) we obtain $\langle J^k \rangle = 0$.
Any other system of gapped noninteracting fermions connected with such a homogeneous system by smooth transformation will have vanishing total electric current.

\subsection{Introduction of interactions between the fermions.}

\label{SectCoulomb}


Now let us take into account interactions \rv{between the fermions. } Our consideration here to some extent repeats the one presented in \cite{ZZ2019_2}. In order to calculate electric current, we consider variation of partition function caused by the  variation of external electromagnetic field $A$. In order to introduce interaction among the fermions, we consider the system with the Euclidean action
\begin{eqnarray}
S&=& \int  dp \bar{\psi}_{p}\hat{Q}(p,i\partial_p)\psi_{p}\nonumber\\&&
+\alpha \int  dp dq dk \bar{\psi}_{p+q}\psi_{p}\tilde{V}({\bf q})\bar{\psi}_{k}\psi_{q+k}
\end{eqnarray}
Here operator $\hat{Q}$ depends on the operators of spatial coordinates $i\partial_p$ because the external field has been included (see Eq.(\ref{Z01})).
For definiteness we consider the Coulomb interaction $V({\bf x})=1/|{\bf x}|=1/\sqrt{x_1^2+x_2^2}$,
for ${\bf x}\not= {\bf 0}$. However, the other types of interactions that occur due to the exchange by bosonic excitations are similar.
Then the Fourier transformed Coulomb potential is
$\tilde{V}({\bf q}) =\sum_{\bf x} \frac{e^{i{\bf q\cdot x}}}{\sqrt{x_1^2+x_2^2}}$.
%
The Coulomb interaction contributes to the self-energy of the fermions,
and the leading order contribution is proportional to $\alpha$.
The Green function can be calculated through the Feynman diagrams as follows
\begin {eqnarray}\label{Green_a}
&& G_{\alpha}(x,y) =  G_{0}(x,y) \nonumber\\&&+ \int G_{0}(x,z_1) \Sigma(z_1,z_2)G_{0}(z_2,y)dz_1 dz_2\nonumber\\&& + \int G_{0}(x,z_1) \Sigma(z_1,z_2)G_{0}(z_2,z_3)\nonumber\\&&\Sigma(z_3,z_4)G_{0}(z_4,y)dz_1 dz_2 dz_3 dz_4 +...
\end{eqnarray}
where
\begin {eqnarray}\label{Sigma_1}
\Sigma(z_1,z_2)&=&\alpha G_{0}(z_1,z_2) \delta(\tau_1-\tau_2)V({\bf z}_1-{\bf z}_2)+O(\alpha^2),\nonumber
\end{eqnarray}
with $z_i=({\bf z}_i,\tau_i)$.
Using Wigner transformation, one finds that
\begin {eqnarray}\label{Green_b_Wigner}
&&G_{\alpha,W}(R,p) = G_{0,W}(R,p)\nonumber\\
&& +  G_{0,W}(R,p)\star  \Sigma_W(R,p)\star  G_{0,W}(R,p) + ...,
\end{eqnarray}
where $G_{0,W}(R,p)$ satisfies $Q_{0,W}(R,p)\star  G_{0, W}(R,p) = 1$, equivalent to Eq.(\ref{GQW}),
while $\Sigma_W$ is Wigner transformation of $\Sigma$.
\rv{It is easy to find that $G_{\alpha,W}(R,p)$ satisfies
\begin {eqnarray}\label{GQW_a}
Q_{\alpha,W}(R,p) \star G_{\alpha,W}(R,p) = 1,
\end{eqnarray}
where $ Q_{\alpha,W}(R,p) =Q_{0,W}(R,p) -\Sigma_W$.
}

\begin{figure}[h]
	\centering  %
	\includegraphics[width=5cm]{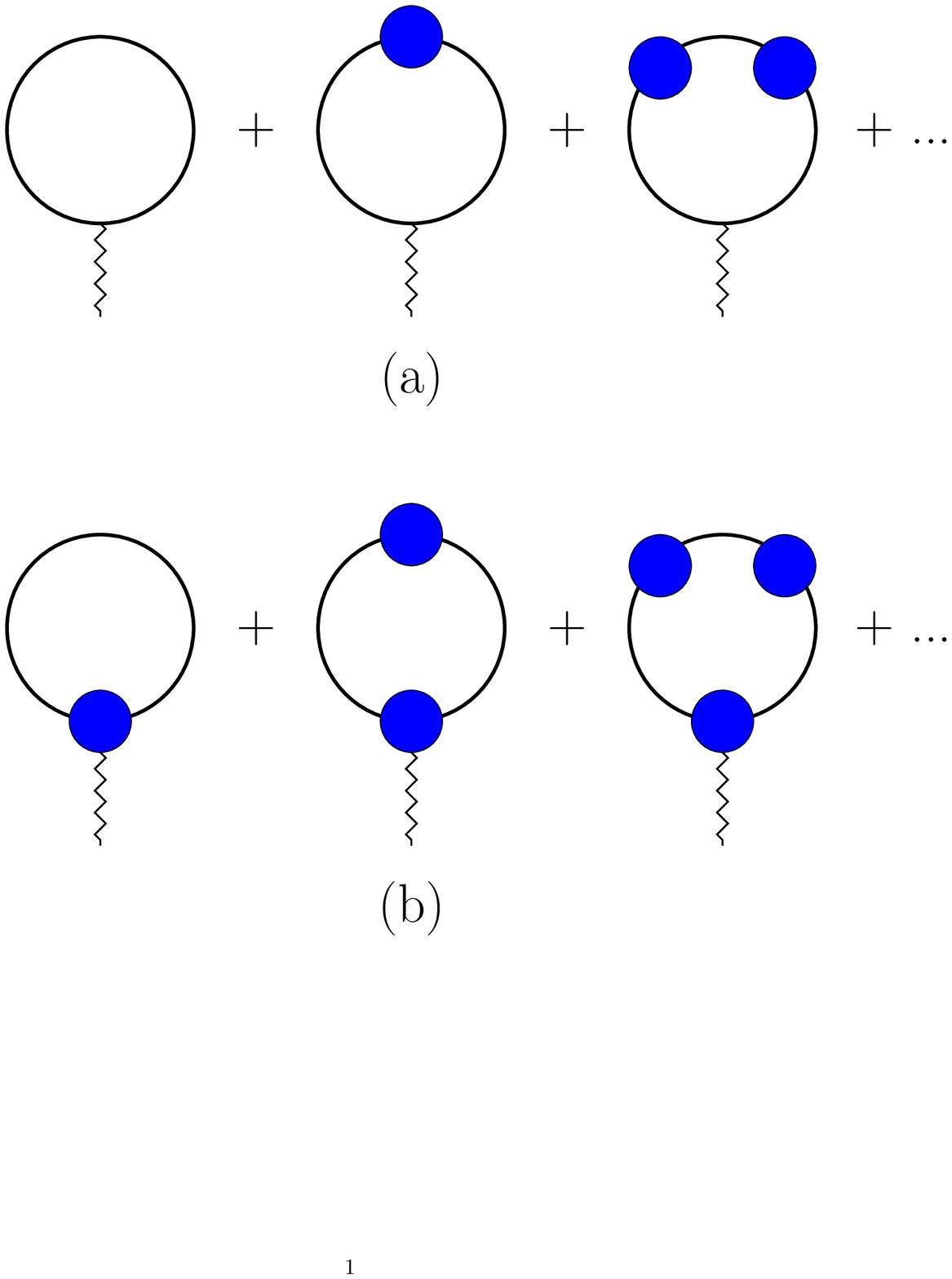}  %
	\caption{a) Feynmann diagrams  for $I(\alpha) = \int \frac{Td^DR}{V} \frac{d^Dp}{(2\pi)^D} Tr G_{\alpha,W} \partial_{p_x} Q_{0,W}$ (expression for the electric current). The filled circles mark $\Sigma_W$. The external wavy line marks the position of $\partial_{p_x} Q_{0,W}$.
         b) Feynmann diagrams for $\Delta I(\alpha) = \int \frac{Td^DR}{V} \frac{d^Dp}{(2\pi)^D} Tr G_{\alpha,W} \partial_{p_x} \Sigma_{W}$. The filled circle with the external wavy line marks $\partial_{p_x} \Sigma_{W}$.}  %
	\label{Fig_tadpole}   %
\end{figure}

Without loss of generality, we can consider only electric current along the $x$-axis,
$J_x$, i.e. $k=1$ in Eq.(\ref{J2}). In what follows we will denote $I=\langle J_x \rangle$.
It is convenient to expand $G_{\alpha,W}(R,p)$ in powers of the coupling constant $\alpha$
as $G_{\alpha,W}= {\cal G}_0 + \alpha {\cal G}_1 +\alpha^2 {\cal G}_2 +... $.
The average total electric current divided by the sustem volume $V$ may also be expanded in powers of $\alpha$:
\begin {eqnarray}
&&I(\alpha) = \frac{T}{V}\int d^D R \int \frac{d^D p}{(2\pi)^D}  \nonumber\\
&& Tr G_{\alpha,W}(R,p) \star  \frac{\partial}{\partial p_x} Q_{0,W}(R,p)\nonumber\\
         &&=  \frac{T}{V} \int d^D R \int \frac{d^D p}{(2\pi)^D}  Tr \big[ G_{0,W} \nonumber\\
         &&+ \sum_{n=1,2,...}G_{0,W}(\star \Sigma_W\star G_{0,W})^n \big]\star  \frac{\partial}{\partial p_x} Q_{0,W}(R,p) \nonumber \\
\end{eqnarray}\label{current_3D_a}
\rv{The corresponding Feynman diagrams are shown in Fig.\ref{Fig_tadpole}(a).}
We represent $\Sigma_W = \alpha {\cal S}_1 +\alpha^2  {\cal S}_2 +... $,
and the current ($x$-component) is given by
 $I=  {\cal I}_0 + \alpha {\cal I}_1 +\alpha^2 {\cal I}_2 +... $,
in which ${\cal I}_0= \frac{T}{V}\int d^D R \int \frac{d^D p}{(2\pi)^D}
Tr G_{0,W}\star \partial_{p_x} Q_{0,W}$,
and
\begin {eqnarray}\label{current_i}
&& {\cal I}_r =\int \frac{Td^D R d^Dp}{V(2\pi)^D} \nonumber \\
&& {\rm Tr} \sum_{l_1+...+l_n=r}\, G_{0,W} \star \prod^n_{i = 1}\Big[  {\cal S}_{l_i}
\star G_{0,W}\star\Big] \frac{\partial}{\partial p_x} {\cal Q}_{0,W},\nonumber \\
\end{eqnarray}
with $r\geq 1$.
Let us compare the obtained expression for the total electric current
with the following expression written through the interacting Green function
\begin{equation}
\tilde{I}(\alpha) = \int \frac{Td^D R d^Dp}{V(2\pi)^D}  Tr G_{\alpha,W}(R,p) \star  \frac{\partial}{\partial p_x} Q_{\alpha,W}(R,p)\label{tildeI}
\end{equation}
For this purpose we calculate the difference
$\Delta I(\alpha) = I(\alpha)-{\tilde I}(\alpha)$.
\rv{Because $Q_{\alpha,W}=Q_{0,W}-\Sigma_W$,}
$\Delta I(\alpha)$ is given by
\begin {eqnarray}  \label{current_3D_a}
\Delta I &=& \int \frac{Td^D R}{V} \int \frac{d^D p}{(2\pi)^D}  Tr G_{\alpha,W}(R,p) \star  \frac{\partial}{\partial p_x} \Sigma_{W}(R,p)\nonumber\\
         &=& \int \frac{Td^D R}{V} \int \frac{d^D p}{(2\pi)^D}  Tr \Big( G_{0,W} \nonumber\\&&+ \sum_{n=1,2,...}G_{0,W}
             (\star \Sigma_W\star G_{0,W})^n \Big)
             \star  \frac{\partial}{\partial p_x} \Sigma_{W}(R,p)\nonumber\\
         &=& \alpha \int \frac{Td^D R}{V} \int \frac{d^D p}{(2\pi)^D}  Tr G_{0,W}\star \frac{\partial}{\partial p_x} {\cal S}_1(R,p)\nonumber\\
        && +\alpha^2 \int \frac{Td^D R}{V} \int \frac{d^D p}{(2\pi)^D} \Big(Tr {\cal S}_1 \star  G_{0,W}\nonumber\\
       &&\star \frac{\partial}{\partial p_x} {\cal S}_1(R,p) \star  G_{0,W} \nonumber\\
        && +  Tr  G_{0,W}\star \frac{\partial}{\partial p_x} {\cal S}_2(R,p) \Big)+...
\end{eqnarray}
The Feynmann diagrams corresponding to $\Delta I$ are represented in Fig. \ref{Fig_tadpole} (b).
Let us consider the diagram with $n$ self-energy functions $\Sigma_W$ (in addition to an extra self-energy with a photon tail)
\begin{eqnarray}
\Delta I^{(n)} =  \int \frac{Td^D R}{V} \frac{d^Dp}{(2\pi)^D} Tr (G_{0,W}\star \Sigma_{W}\star )^n G_{0,W}\star  \partial_{p_x} \Sigma_{W},
\end{eqnarray}
which appeared in the third line in Eq.(\ref{current_3D_a}). After partial integration, we obtain
\begin{eqnarray}
&&\Delta I^{(n)} =
 (n+1)\int \frac{Td^D R d^Dp}{S(2\pi)^D}\nonumber\\&& {\rm Tr} G_{0,W}\star \partial_{p_x} Q_{0,W}\star  G_{0,W}...\star  \Sigma_{W} \nonumber\\
 &&-n\int \frac{Td^D R d^Dp}{S(2\pi)^D} Tr G_{0,W}\star \partial_{p_x}\Sigma_W\star \nonumber\\
&&...\star \Sigma_{W}\star G_{0,W}\star  \Sigma_{W}\nonumber
\end{eqnarray}
We come to the following relation
\begin{eqnarray}
&&(n+1)\Delta I^{(n)} = (n+1)\int \frac{Td^D R}{V} \frac{d^Dp}{(2\pi)^D}\nonumber\\
&& {\rm Tr} G_{0,W}\star \partial_{p_x} Q_{0,W}\nonumber\\
&&\star  G_{0,W}\star ...\star \Sigma_{W}\star G_{0,W}\star  \Sigma_{W},
\end{eqnarray}
which gives
$
\Delta I^{(n)} = I^{(n+1)}
$,
where $I^{(n+1)}$ is the contribution to electric current with $n+1$ insertions of $\Sigma_W$
represented schematically in Fig. \ref{Fig_tadpole} (a) (the \rv{ $n+1$ -th } term in the sum).
Overall, we obtain:
$$
\Delta I(\alpha) = I(\alpha)-I^{(0)} = I(\alpha)-I(0)
$$
We find that the total current is given by an integral of Eq. (\ref{tildeI}) as long as the value of the total current remains equal to its value without interactions. We will prove that indeed $I(\alpha) = I(0)$ in the region of analyticity in $\alpha$, i.e. as long as the perturbation theory in $\alpha$ may be used.

\begin{figure}
			\includegraphics[width=5cm]{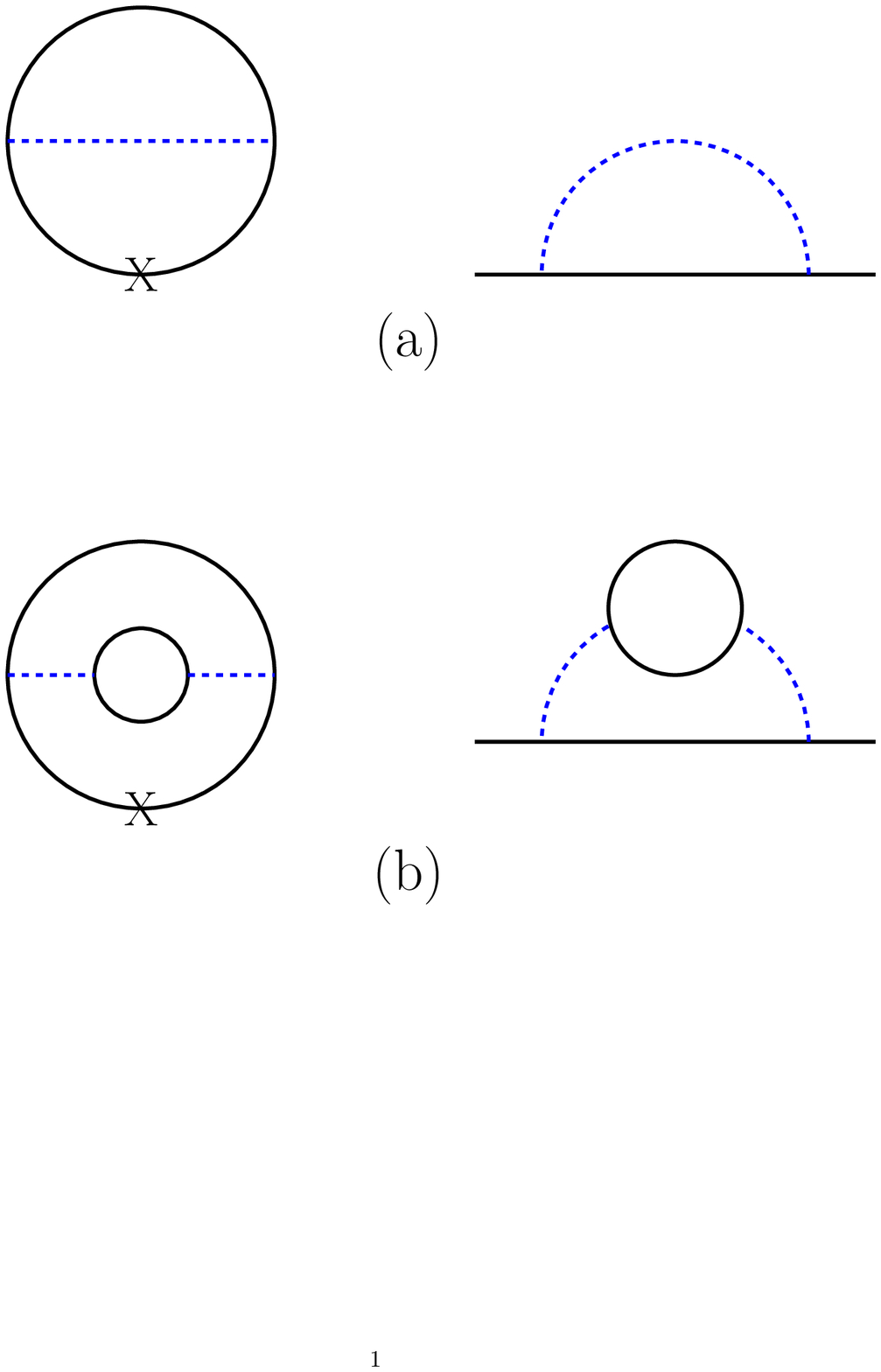}
	\caption{  Loop diagram contributions to electric
current $I$ (left side of the figure), and the corresponding diagrams
of self-energy function (right side). \rv{Crosses "X" represent $\partial_{p_x} Q_{0,W}$.}
(a) The diagrams in the first order.
(b) One of the second order diagrams.
}
	\label{sunrise}
\end{figure}
%


The electric current in the absence of interactions is given by
${\cal I}_0 =  \int \frac{Td^D R}{V} \int \frac{d^D p}{(2\pi)^D}
Tr G_{0,W}(R,p) \star  \frac{\partial}{\partial p_x} Q_{0,W}(R,p)$.
Below we will prove that this expression does not receive corrections from interactions,
i.e. for  $j \geq 1$, ${\cal I}_j=0$.
\begin{figure}
			\includegraphics[width=5cm]{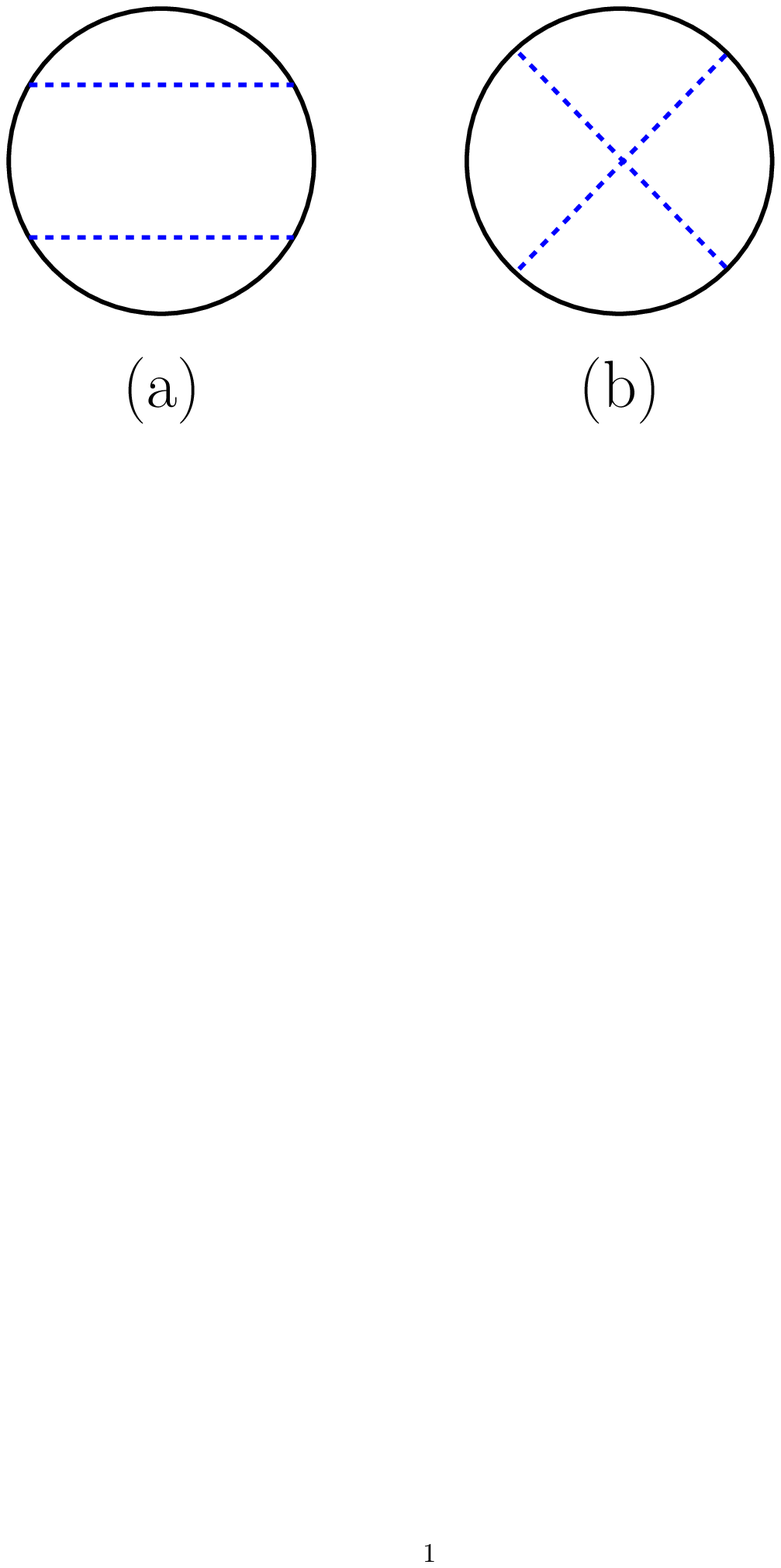}
	\caption{ a) The progenitor diagram for the two - loop rainbow contribution to electric current. b) The progenitor diagram for the two - loop contribution to electric current (which is {beyond the} rainbow approximation). 
}
	\label{progenitor}
\end{figure}
First, let us consider ${\cal I}_1$ \rv{ (shown in Fig.\ref{sunrise}(a)), }
which can be expressed explicitly as
\begin {eqnarray}\label{current_1st}
&&{\cal I}_1 =-\int \frac{Td^D R}{V} \int \frac{d^D p d^D q}{(2\pi)^D} \nonumber\\
&&{\rm Tr}  ( G_{0,W}(R,p-q){\cal D}(q))
\star  \frac{\partial}{\partial p_x} G_{0,W}(R,p) \nonumber\\
&&=-\int \frac{Td^D R}{V} \int \frac{d^D p d^D q}{(2\pi)^D}\nonumber\\
&&{\rm  Tr}  ( G_{0,W}(R,p-q){\cal D}(q) )\nonumber\\
&&  \frac{\partial}{\partial p_x} G_{0,W}(R,p)
\end{eqnarray}
Here ${\cal D}(q)$ is the Fourier transformation of function
\begin {eqnarray}\label{Sigma_1}
D(z_1-z_2)&=&\alpha \delta(\tau_1-\tau_2)V({\bf z}_1-{\bf z}_2).\nonumber
\end{eqnarray}
Because ${\cal D}(q)$ is an even function,
for each value of $R$ the above expression is proportional to
\begin {eqnarray}\label{lemma_1}
\int\int {\cal F}_R(p-q){\cal D}_R(q){\cal F}_R'(p)dpdq=0,
\end{eqnarray}
where  ${\cal F}_R(q) = G_{0,W}(R,q)$, and ${\cal F}'$ is the first derivative of $\cal F$.
This representation allows us to prove that ${\cal I}_1 =0$
(we perform the integration by parts and show that ${\cal I}_1 = -{\cal I}_1$).
\begin{figure}[h]
	\centering  %
	\includegraphics[width=5cm]{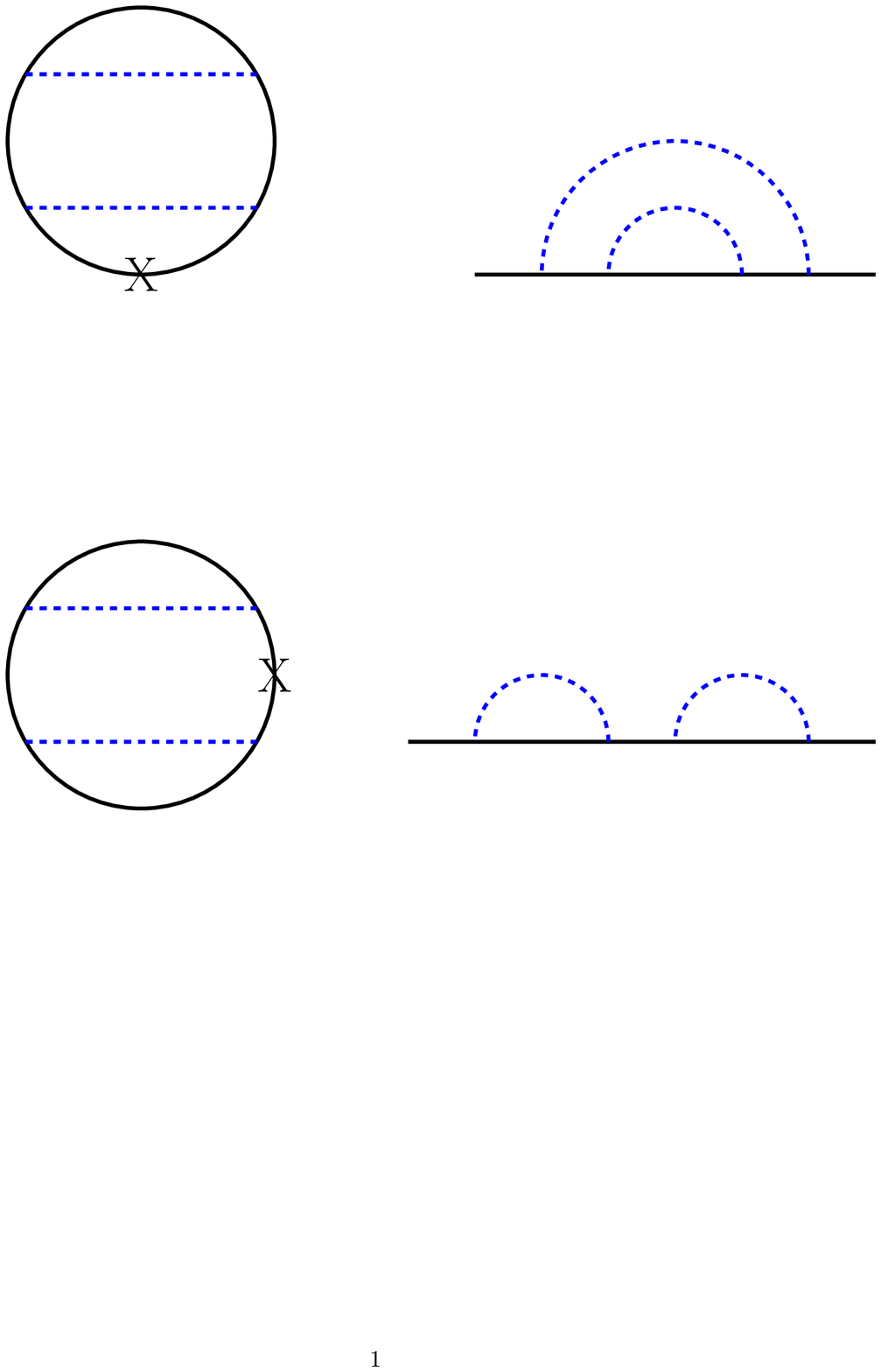} \vspace{1cm} %
	\caption{Two loop Feynmann diagrams for the self energy $\Sigma$ in rainbow approximation (right side of the figure) and the corresponding three loop rainbow contributions to electric current $I$ (left side of the figure). The crosses point out the positions of the derivatives $\partial_{p_x}Q_{0,W}$.}  %
	\label{fig_rainbow}   %
\end{figure}

\begin{figure}[h]
	\centering  %
	\includegraphics[width=5cm]{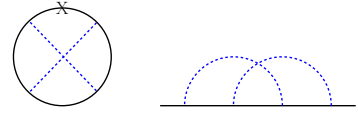}\vspace{1cm}  %
	\caption{Two loop Feynmann diagrams for the self energy $\Sigma$ {beyond the} rainbow approximation (right side of the figure), and the corresponding three loop contributions to electric current $I$ (left side of the figure). The crosses point out the positions of the derivatives $\partial_{p_x}Q_{0,W}$.}  %
	\label{fig_cross}   %
\end{figure}

Let us now consider the next order contribution ${\cal I}_2$. We have
\begin{eqnarray}\label{current_i2}
&&{\cal I}_2 =-\int \frac{Td^D R d^Dp}{V(2\pi)^D} Tr {\cal S}_2 \star
               \frac{\partial}{\partial p_x} G_{0,W}\nonumber \\&&
-\int \frac{Td^D R d^Dp}{V(2\pi)^D}
 Tr {\cal S}_1 \star  G_{0,W} \star  {\cal S}_1\star  \frac{\partial}{\partial p_x} G_{0,W}\nonumber
\end{eqnarray}
\rv{First, similar to the proof of ${\cal I}_1=0$,
the contribution of the diagram shown in Fig.\ref{sunrise}(b) is also zero.
The only necessary change in the proof is to replace ${\cal D}(q)$ in Eq.(\ref{current_1st})
 by ${\cal D}(q)\Pi(q^2){\cal D}(q)$, where $\Pi(q^2)$ is the vacuum polarization.}
Taking self-energy ${\cal S}_2$ in rainbow(r.b.) approximation
(shown in the right side of Fig.\ref{fig_rainbow}), we get 
\begin{eqnarray}\label{current_i2}
 &&{\cal I}^{(r.b.)}_2 =
-\int  \frac{T d^D R d^D pd^D k d^D q}{V(2\pi)^{2D}} \,Tr \Big[G_{0,W}(R,p-k)\nonumber\\
&&\star G_{0,W}(R,p-k-q){\cal D}(q)\star  G_{0,W}(R,p-k)\Big]\nonumber\\
&&{\cal D}(k)\star \partial_{p_x}G_{0,W}(R,p)  \nonumber\\
&& -\int \frac{T d^D Rd^D pd^D k d^D q}{V(2\pi)^{2D}}\,Tr G_{0,W}(R,p-q){\cal D}(q)\nonumber\\
&&\star  G_{0,W}(R,p)\star G_{0,W}(R,p-k){\cal D}(k)\star \partial_{p_x}G_{0,W}(R,p)\nonumber
\end{eqnarray}
In the first term the star before $\partial_{p_x}$ may be eliminated.
It may then be inserted before ${\cal D}(k)$, thus giving
\begin{eqnarray}\label{current_i2}
&&{\cal I}^{(r.b.)}_2 =
-\int \frac{T d^D Rd^D pd^D k d^D q}{V(2\pi)^{2D}} \,Tr \Big[G_{0,W}(R,p-k)\nonumber\\
&&\star G_{0,W}(R,p-k-q){\cal D}(q)\star  G_{0,W}(R,p-k)\Big]\star \nonumber\\
&&{\cal D}(k)\partial_{p_x}G_{0,W}(R,p)  \nonumber\\
&& -\int \frac{T d^D Rd^D p}{V(2\pi)^D} \,Tr G_{0,W}(R,p-q){\cal D}(q)\nonumber\\
&&\star  G_{0,W}(R,p)\star G_{0,W}(R,p-k){\cal D}(k)\nonumber\\
&&\star \partial_{p_x}G_{0,W}(R,p)\nonumber\\
&=&
 -\frac{1}{2}\int \frac{T d^D Rd^D pd^D k d^D q}{V(2\pi)^{2D}}\,\partial_{p_x} \,Tr \Big[G_{0,W}(R,p-k)\nonumber\\
 &&\star G_{0,W}(R,p-k-q){\cal D}(q)\star  G_{0,W}(R,p-k)\Big]\star \nonumber\\
 &&{\cal D}(k)G_{0,W}(R,p) = 0
\end{eqnarray}
Notice, that the last expression without derivative with respect to $p_x$ corresponds to the diagram
similar somehow to the one called in \cite{parity_anomaly} "progenitor". We present the form of
the corresponding Feynmann diagram in Fig. \ref{progenitor}(a) and call it the progenitor
for the diagrams presented in Fig. \ref{Fig_tadpole}. In essence, our present proof is an extension of the one given in \cite{parity_anomaly}.
The remaining two loop diagrams (see Fig. \ref{fig_cross}) give the contribution that may be written as follows
\begin{eqnarray}\label{current_i2}
&& {\cal I}^{(cross)}_2 = -\int  \,Tr \Big[ G_{0,W}(R,p-k_1)\star \nonumber\\
&& G_{0,W}(R,p-k_1-k_2)\star  G_{0,W}(R,p-k_2)\Big]\nonumber\\
&&{\cal D}(k_1) {\cal D}(k_2)\nonumber\\
&&\partial_{p_x}G_{0,W}(R,p)\frac{T d^D Rd^D pd^D k_1 d^D k_2}{V(2\pi)^{2D}} \nonumber\\
&&=-\frac{1}{4}\int \frac{T d^D Rd^D pd^D k d^D q}{V(2\pi)^{2D}}\partial_{p_x} \,Tr \Big[ G_{0,W}(R,p-k_1)\star \nonumber\\
&& G_{0,W}(R,p-k_1-k_2)\star  G_{0,W}(R,p-k_2)  \nonumber\\
&&\star G_{0,W}(R,p)\Big]{\cal D}(k_1) {\cal D}(k_2) = 0
\end{eqnarray}
Here the star $\star =e^{i\overleftarrow{\partial}_R\overrightarrow{\partial}_p/2-i\overleftarrow{\partial}_p\overrightarrow{\partial}_R/2}$
acts only on $G$, but does not act on ${\cal D}(k_i)$ (which doesn't depend on $R$ or $p$).
The last line of the above expression corresponds to the diagram of Fig. \ref{progenitor} (b).

One can see, that  ${\cal I}_2=0$. In the same way the higher orders may be considered. 
One can check that ${\cal I}_j=0$ for $j > 0$ to all orders of the perturbation theory.

\begin{figure}[h]
	\centering  %
	\includegraphics[width=5cm]{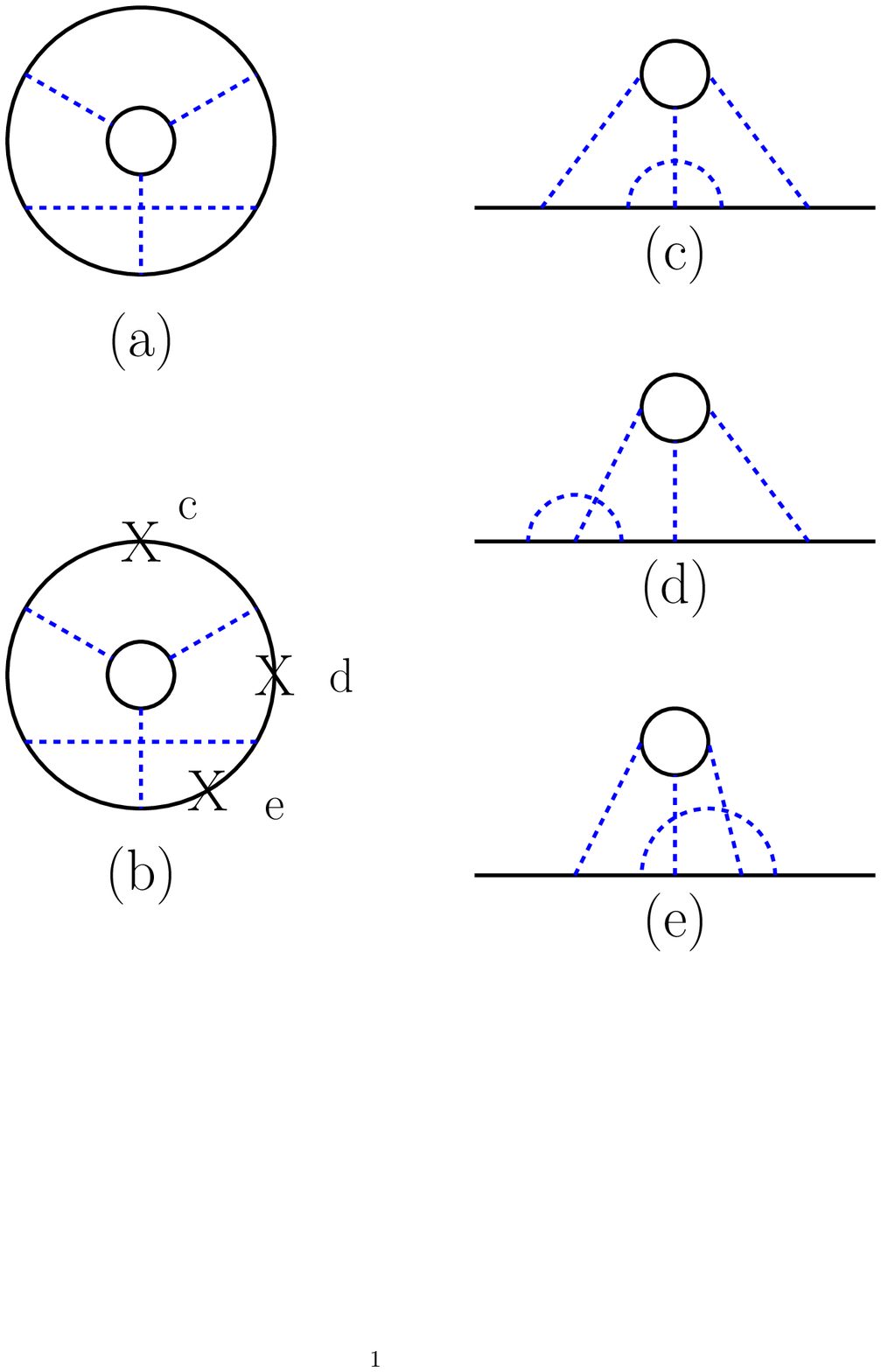}\vspace{1cm}  %
	\caption{An example of the high-order corrections.
(a) is the progenitor diagram, the possible contributions to the electric current appear when in (b) one of the crosses is substituted by the derivative $\partial_p Q_{0,W}$.
Using each of those crosses we form the diagram, which contributes to electric current.
Diagrams (d), (e), and (f) are the corresponding self-energy diagrams.
 }  %
	\label{fig_high-order}   %
\end{figure}

The higher order corrections may be considered in the similar way. The example of the higher order diagram is considered in Fig.\ref{fig_high-order}. The sum of the Feynman diagrams represented in Fig.\ref{fig_high-order} (c), (d), (e) contribute the Fermion self energy that enters an expression for the total current presented in Fig. \ref{Fig_tadpole} (a) (the diagrams (d) and (e) are to be counted twice). The resulting contribution to electric current is equal to the integral over momentum of the derivative of the progenitor diagram represented in Fig.\ref{fig_high-order} (a). This integral is zero for the system with  compact momentum space (when lattice regularization is used). The diagrams of Fig.\ref{fig_high-order} (c), (d), (e) appear when the diagram of  Fig.\ref{fig_high-order} (b) is cut at the positions of the crosses.

The obtained results mean the following:
(1) The interaction corrections to the total electric current vanish.
(2) There is the following representation for the total average electric current divided by the system volume in the considered system:
\begin {equation}
{I}(\alpha) = \int \frac{Td^D R d^Dp}{V(2\pi)^D} Tr G_{\alpha,W}(R,p) \star  \frac{\partial}{\partial p_x} Q_{\alpha,W}(R,p)\label{IFIN}
\end{equation}

Notice that our proof doesn't rely on the precise expression of Coulomb potential.
In Eq.(\ref{lemma_1}) we only used that the Fourier-transformed potential is even function of momentum. Therefore, the generalization of  our result to the case of the other interactions is straightforward. In the similar way the Yukawa interaction, the exchange by gauge bosons and the four-fermion interactions may be considered.

\section{Gapless fermions}

\subsection{Electric current in the system of gapless noninteracting charged fermions}

In this section we discuss the case of the gapless fermions. Let us start from the consideration of the noninteracting fermions with Hamiltonian $\hat{H}$. Then
$$
\hat{Q} = i\omega - \hat{H}
$$
For the case of the homogeneous system with $\hat{H} = H(\hat{p})$ Eq. \ref{J2} receives the form
\begin{eqnarray}
\langle J^k \rangle &=&  V \, \int \frac{d\omega d^{D-1} p}{(2\pi)^D}\, {\rm Tr} \, \frac{1}{i\omega - H(p)}  \partial_{p_k} H(p)  \label{J2_}
\end{eqnarray}
As it was mentioned above the integral over $\omega$ is to be regularized if we are interested in the expression for the current out of the linear response to external gauge field. We modify $G_W(p,x) \to G_W^{(reg)}(p,x) = e^{i\tau \omega} G_W(p,x)$, where $\tau$ be set to $\tau \to +0$ at the end of calculations. The integral $\int d\omega \frac{e^{i\omega \tau}}{i\omega - {\cal E}_n}$ entering Eq. (\ref{J2_}) (here ${\cal E}_n$ is the $n$ - th eigenvalue of the Hamiltonian) may be calculated using the residue theorem:
$$
\int d\omega \frac{e^{i\omega \tau}}{i\omega - {\cal E}_n}\Big|_{\tau = +0} = 2\pi \theta(-{\cal E}_n)
$$
provided that ${\cal E}_n \ne 0$. In the case when  the value of ${\cal E}_n$ vanishes, this integral is divergent. This breaks the topological nature of Eq. (\ref{J2_}) but does not mean that the whole Eq. (\ref{J2_}) is divergent itself. Namely, we have
\begin{eqnarray}
\langle J^k \rangle &=&  V \, \int \frac{ d^{D-1} p}{(2\pi)^{D-1}}\, {\rm Tr} \,  \theta(- H(p))  \partial_{p_k} H(p)  \label{J2__}
\end{eqnarray}
For the case of the one - dimensional ($D=2$) system with one branch of spectrum we get
\begin{eqnarray}
\langle J \rangle &=&  \frac{V}{2\pi} \, \int_{p_1}^{p_2} dp \,  \partial_{p} H(p) = \frac{V}{2\pi} (H(p_2)  - H(p_1))   \label{J2__}
\end{eqnarray}
where $p_1$ and $p_2$ are the endpoints of the piece of the  branch of spectrum with $H(p)\le 0$. If both $p_1$ and $p_2$ are finite,
\rev{ then $H(p_1)=H(p_2)=0$, therefore $\langle J^k \rangle =0$.} This occurs for the compact Brillouin zone that appears for the lattice tight - binding model. If one of the points $p_1$ and $p_2$ are placed at infinity, then the value of the total current may differ from zero. In this case this is clear that this expression depends continuously on the smooth variation of the Fermi energy. 

Thus we are able to give another weakened version of Bloch theorem valid for the gapless systems: the homogeneous lattice model of noninteracting fermions cannot have the non - vanishing total electric current.
Since we cannot formulate at the present moment a  more general version of Bloch theorem for the gapless QFT system, we consider below several particular examples. Some of them break the conventional Bloch theorem.

\subsection{An example of the system that obeys Bloch theorem}

\label{SectE1}

In this section we consider the planar system placed in the $(xy)$ plane: in the
region $x<0$ there is an infinitely high potential, while in the region $x>0$ there is the constant electric field directed towards the
positive $x$-axis, and uniform magnetic orthogonal to the  $(x,y)$ plane.

Electron in such a system satisfies the following Schrodinger equation
\begin{eqnarray}\label{sch_equ_1}
-\frac{1}{2m}\partial^2_x \psi+\frac{(-i\partial_y-Bx)^2}{2m} \psi+V(x)\psi=\epsilon\psi
\end{eqnarray}
where $V(x)=Ex$ for $x>0$. Notice that we use the relativistic system of units with $\hbar = c = 1$.
Separating variables $\psi(x,y)=e^{ip_y y}\phi(x)$, one obtains the equation for $\phi(x)$ as follows
\begin{eqnarray}\label{sch_equ_2}
&&\phi''(x)- B^2 \Big(x-\frac{p_y B-mE}{B^2}\Big)^2\phi(x)
\nonumber\\&&+\Big({2m\epsilon}{}-\frac{2m p_y E}{ B}+\frac{m^2 E^2}{B^2}\Big)\phi(x)=0.
\end{eqnarray}
We rescale variable $x=\kappa s$, with $\kappa=1/\sqrt{2B}$, and arrive at
\begin{eqnarray}\label{sch_equ_3}
f''(s)-\frac{1}{4}(s - s_0)^2 f(s)+(\nu+\frac{1}{2}) f(s)=0,
\end{eqnarray}
where $s_0=\sqrt{2/B}p_y -mE\sqrt{2B}/B^2$, and
\begin{eqnarray}\label{energy}
\nu+\frac{1}{2}
=\frac{m\epsilon}{B}-\frac{m p_y E}{ B^2}+\frac{m^2 E^2}{2 B^3}.
\end{eqnarray}

\begin{figure}[h]
	\centering  %
	\includegraphics[width=5cm]{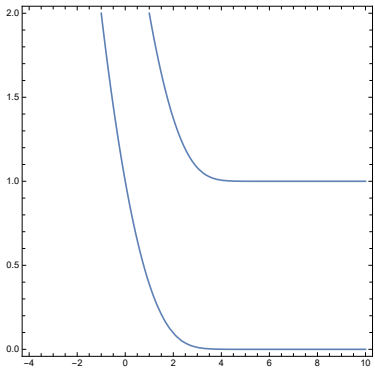}  %
	\caption{$\nu$ versus $s_0$ in the model of Sect. \ref{SectE1}.}
	\label{fig_Spec_a}   %
\end{figure}

Solution of Eq.(\ref{sch_equ_3}) with the requirement $s\rightarrow\infty, f\rightarrow 0$
is the  parabolic cylinder function $P(\nu,s-s_0)$. From the boundary condition $f(s=0)=0$
we obtain relation between $\nu$ and $s_0$ (see Fig.  \ref{fig_Spec_a}).

\begin{figure}[h]
	\centering  %
	\includegraphics[width=5cm]{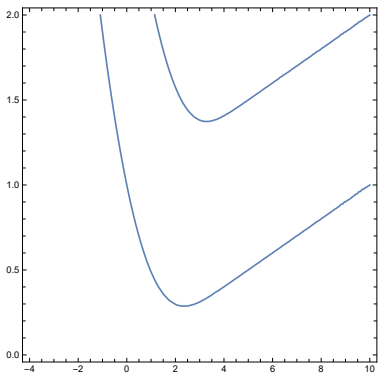}  %
	\caption{Relation between energy and momentum in the model of Sect. \ref{SectE1}.
              Vertical axis is $m\epsilon / ( B)$, while the horizontal axis is $s_0$. We take the particular case of $mE/B^2=0.1$.}
	\label{fig_Spec_b}   %
\end{figure}

$\nu$ linearly depends on energy $\epsilon$ and momentun $p_y$,
while $s_0$ linearly depends on momentum $p_y$. Finally, we obtain
relation between $\epsilon$ and $p_y$, which is shown in Fig. \ref{fig_Spec_b}.

The total current is equal to
$$
j_y = \sum_{i=1,2}\int^{p^{(right)}}_{p^{(left)}} \frac{d p_y}{2\pi} \partial_{p_y}  {\cal E}_{i}(p_y)
$$
where the integral is between the two crossing points $p_y=p^{(left)},p^{(right)}$ (of the Fermi level and the given branch of spectrum). One can see that the total current is equal to
zero.  Therefore, the Bloch theorem is valid
in this case.

\subsection{A counter-example}

\label{SectE2}

Now let us consider another example.
This is an infinite planar system in the $xy$ plane
with magnetic field $B$ penetrating the plane: in the region
 $x<0$ there is a uniform magnetic field $B_z=-B <0$,
while in the region $x>0$, there is a uniform magnetic field in the opposite direction, i.e. $B_z= B >0$.

\begin{figure}[h]
	\centering  %
	\includegraphics[width=5cm]{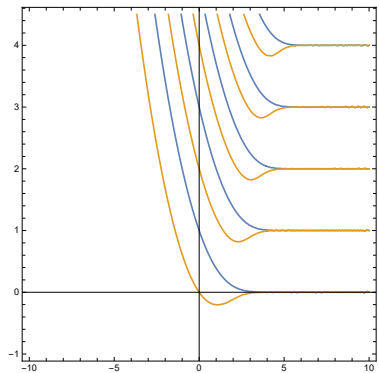}  %
	\caption{$\nu$ as a function of $s_0$ in the model of Sect. \ref{SectE2}. The blue lines come from the condition $P(\nu,-s_0)=0$,
                  while the brown lines from the condition $Q(\nu,-s_0)=0$. }
	\label{fig_Spec_2mag_a}   %
\end{figure}

An electron  in such a system satisfies the following Schrodinger equation
\begin{eqnarray}\label{sch_equ_a1}
-\frac{1}{2m}\partial^2_x \psi+\frac{(-i\partial_y- Bx)^2}{2m} \psi+V(x)\psi=\epsilon\psi , x>0 \\
-\frac{1}{2m}\partial^2_x \psi+\frac{(-i\partial_y+Bx)^2}{2m} \psi+V(x)\psi=\epsilon\psi ,   x<0
\end{eqnarray}
After separation of variables and rescaling $x$ via $x=\kappa s$, one obtains
\begin{eqnarray}\label{sch_equ_a2}
f''(s)-\frac{1}{4}(s - s_0)^2 f(s)+(\nu+\frac{1}{2}) f(s)=0,
\end{eqnarray}
for $s>0$, where $s_0=\sqrt{2/B} p_y$ and
$$\nu+1/2=m\epsilon/B.$$
The equation for the region $s<0$ is similar; the only difference is a sign change  in front of $s$ in Eq.(\ref{sch_equ_a2}).
The solution can be expressed in terms of parabolic cylinder function:
\begin{equation}
f(s)=
\left\{
 \begin{array}{lr}
C_1 P(\nu,s-s_0) &  x>0,  \\
C_2 P(\nu,-s-s_0) & x<0.\\
 \end{array}\right.
\end{equation}
The boundary condition is that $f(s)$ and $f'(s)$ should be continuous at $s=0$.
If we denote the derivative function of $P(\nu,s)$ respect to $x$ as $Q(\nu,s)$,
the boundary condition implies $P(\nu,-s_0)=0 $ (then $C_1 = - C_2$)
or $Q(\nu,-s_0)=0 $ (then $C_1 = C_2$).
We find the energy spectrum, which is shown in Fig. \ref{fig_Spec_2mag_a}.

From this spectrum, one can see that if the Fermi level is between $( B/m)(0+1/2)$ and $( B/m)(1+1/2)$, it crosses both branches of spectrum ${\cal E}_{1,2}(p_y)$ corresponding to blue and brown lines in Fig. \ref{fig_Spec_2mag_a}. The total current is equal to
$$
j_y = \sum_{i=1,2}\int^\infty_{p^{(i)}} \frac{d p_y}{2\pi} \partial_{p_y}  {\cal E}_{i}(p_y)
$$
where the integral is between the crossing point $p_y=p^{(i)}$ (of the Fermi level and the given branch of spectrum) and $p_y = +\infty$. One can see that the total current is nonzero.  Therefore, the Bloch theorem is violated in this case.

\subsection{The system with magnetic field in the quantum well}

\label{SectE3}

Now let us consider the more realistic example.
This is an infinite planar system in the $xy$ plane
with magnetic field $B$ penetrating the plane
as in the previous section:
in the region
$x<0$ there is a uniform magnetic field $B_z=-B <0$,
while in the region $x>0$, there is a uniform magnetic field in the opposite direction, i.e. $B_z= B >0$.
Besides, we add the potential $V(x)$ of the quantum well:
$V(x)=0$ for $x\in[-L,+L]$,
and $V(x)=V_0$ for $x\in (\infty, -L)\cup (+L,+\infty)$.

\begin{figure}[h]
	\centering  %
	\includegraphics[width=5cm]{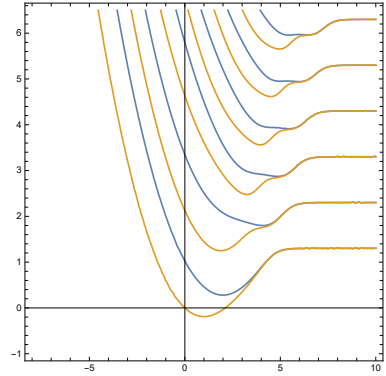}  %
	\caption{$\nu$ as a function of $s_0$ in the model of Sect. \ref{SectE3}. The parameters of the model are $l=3$ and $m V_0/B=1.3$.
                  The blue lines come from the condition $Det^{(1)}=0$,
                  while the brown lines come from the condition $Det^{(2)}=0$. }
	\label{fig_Spec_2mag_b}   %
\end{figure}

An electron  in such a system satisfies the following Schrodinger equation
\begin{eqnarray}\label{sch_equ_b1}
-\frac{1}{2m}\partial^2_x \psi+\frac{(-i\partial_y- Bx)^2}{2m} \psi+V(x)\psi=\epsilon\psi , x>0 \\
-\frac{1}{2m}\partial^2_x \psi+\frac{(-i\partial_y+Bx)^2}{2m} \psi+V(x)\psi=\epsilon\psi ,   x<0
\end{eqnarray}
After separation of variables and rescaling $x$ via $x=\kappa s$, one obtains
\begin{eqnarray}\label{sch_equ_b2}
f''(s)-\frac{1}{4}(s - s_0)^2 f(s)+(\nu+\frac{1}{2}) f(s)=0,
\end{eqnarray}
for $0<s<l=L/\kappa$, where $s_0=\sqrt{2/B} p_y$ and
$$\nu+1/2=m\epsilon/B.$$
When $s>l$, $f(s)$ satisfies
\begin{eqnarray}\label{sch_equ_b3}
f''(s)-\frac{1}{4}(s - s_0)^2 f(s)+(\nu'+\frac{1}{2}) f(s)=0,
\end{eqnarray}
where $\nu'=\nu-m V_0/B$
The equation for the region $s<0$ is similar; the only difference is a sign change
in front of $s$ in Eq.(\ref{sch_equ_b2}) and Eq.(\ref{sch_equ_b3}).
The solution can be expressed in terms of parabolic cylinder function and the hypergeometric function:
\begin{equation}
f(s)=
\left\{
 \begin{array}{lr}
 C_1 F_1(\nu,s-s_0) +  C_2 F_2(\nu,s-s_0) ,      0<x<L;\\
C_3 D_{\nu}(s-s_0),  x>L;  \\
 C'_1 F_1(\nu,-s-s_0) +  C'_2 F_2(\nu,-s-s_0), \\ \quad -L<x<0;\\
C'_3 D_{\nu}(-s-s_0), x<-L.\\
 \end{array}\right.
\end{equation}
$F_1$ and $F_2$ are given by
$F_1(\nu,x)=e^{-x^2/4}F(-\nu/2,1/2;x^2/2)$ and
$F_2(\nu,x)=x e^{-x^2/4} F(1/2-\nu/2,3/2;x^2/2)$.
According to the boundary conditions $f(s)$ and $f'(s)$ are continuous at $s=0,\pm L$,
which leads to 6 linear equations. The corresponding determinant  ($6 \times 6$)
should be zero, which guarantees the nonzero solutions for $C_i$ and $C'_i$.
After linear transformations, the determinant can be decomposed into
the product of two $3 \times 3$ determinants:
\rev{
$Det=2 \cdot Det^{(1)}\cdot Det^{(2)}$, with
\begin{equation}
  Det^{(1)}=
  \begin{vmatrix}
  \mathcal{A} & \mathcal{B} & -\mathcal{P} \\
  \mathcal{F} & \mathcal{G} & -\mathcal{Q} \\
  \mathcal{R}_1 & \mathcal{R}_2 & 0
  \end{vmatrix}
\quad  Det^{(2)}=
  \begin{vmatrix}
  \mathcal{A} & \mathcal{B} & -\mathcal{P} \\
  \mathcal{F} & \mathcal{G} & -\mathcal{Q} \\
  \mathcal{S}_1 & \mathcal{S}_2 & 0
  \end{vmatrix},
\end{equation}
}%
where $\mathcal{A}=F_1(\nu,l-s_0)$, $\mathcal{B}=F_2(\nu,l-s_0)$,
$\mathcal{R}_1=F_1(\nu,-s_0)$, $\mathcal{R}_2=F_2(\nu,-s_0)$,
$\mathcal{P}=D_{\nu'}(l-s_0)$ and $\mathcal{Q}=D'_{\nu'}(l-s_0)$.
The derivatives of $F_1(\nu,x)$ and $F_2(\nu,x)$ respect to $x$ are denoted by  $g_1(\nu,x)$ and $g_2(\nu,x)$, and then
$\mathcal{F}=g_1(\nu,l-s_0)$, $\mathcal{G}=g_2(\nu,l-s_0)$,
$\mathcal{S}_1=g_1(\nu,-s_0)$, $\mathcal{S}_2=g_2(\nu,-s_0)$.
$Det=0$ is equivalent to
$Det^{(1)}=0$  or $Det^{(2)}=0$
from which we find the energy spectrum, i.e.
the relation between $\nu$ and $s_0$,
shown in Fig.\ref{fig_Spec_2mag_b}.

From this spectrum, one can see that if the Fermi level is between $V_0+( B/m)(0+1/2)$ and
$V_0+( B/m)(1+1/2)$, it crosses both branches of spectrum
${\cal E}_{1,2}(p_y)$ corresponding to blue and brown lines in Fig. \ref{fig_Spec_2mag_b}. The total current is equal to
$$
j_y = \sum_{i=1,2}\int^\infty_{p^{(i)}} \frac{d p_y}{2\pi} \partial_{p_y}  {\cal E}_{i}(p_y)
$$
where the integral is between the crossing point $p_y=p^{(i)}$ (of the Fermi level and the given branch of spectrum) and $p_y = +\infty$. One can see that the total current is nonzero.  Therefore, the Bloch theorem is violated in this case.

\section{Conclusions}

\label{SectConcl}

In the present notes we consider the possibility to formulate the analogue of the quantum mechanical Bloch theorem for the field theoretical systems. In the non - relativistic quantum mechanics of fixed number of particles the total current vanishes in equilibrium according to the conventional Bloch theorem. The essential difference from the quantum field theory is that in the latter the number of (quasi)particles is not fixed while the single particle Hamiltonian may have the more complicated form. Moreover, the interactions with the time delay complicate the system even more.
As a result the direct analogue of the Bloch theorem in the QFT has not been established despite several attempts \cite{B2,B3,B4,B5,B6,Watanabe}.

We consider separately the gapped and the gapless systems.  Below we list the obtained results.

\begin{enumerate}

\item{}
First of all, we demonstrate that for the gapped homogeneous noninteracting system with compact Brillouin zone the total electric current vanishes.

\item{}
Next, we prove, that the total electric current for the gapped noninteracting system is the topological invariant in the presence of periodical spatial boundary conditions, i.e. it is not changed when the system is modified smoothly. Therefore, any non - homogeneous smooth modifications of the system mentioned above in item 1 also leads to vanishing total electric current.

\item{}
Interactions due to exchange by bosonic excitations do not alter the total electric current for the mentioned above gapped systems as long as the interactions may be taken perturbatively. We prove this statement to all orders in the coupling constant.

\item{}
Considering the gapless systems we find that the total electric current vanishes for the homogeneous ones with compact Brillouin zone in the absence of interactions.

\item{}
We do not formulate any analogues of the Bloch theorem for the gapless non - homogeneous systems. Instead we consider several particular examples. Along with the ones, where the total current vanishes in equilibrium, we present examples, where the total electric current is nonzero. In those examples space is divided into the pieces with different directions of magnetic field. The total current appears along the interphase between the two pieces. Notice, that this setup does not satisfy conditions of the version of Bloch theorem proposed in \cite{Watanabe,VolovikBloch}. Namely, the considered system is infinite in the direction orthogonal to the persistent equilibrium current.

\end{enumerate}

We conclude, that the Bloch theorem in its traditional formulation ("there is no total electric current in equilibrium") does not hold in quantum field theory. The examples that demonstrate this are those with gapless noninteracting fermions. At the same time we formulate the weakened version of the Bloch theorem for the gapped interacting systems (items 1,2,3 above).

M.A.Z. is indebted for the discussions to G.E.Volovik, who brought to his attention the importance of the possible extension of Bloch theorem to quantum field theory. Both authors  kindly acknowledge discussions with I.Fialkonsky and M.Suleymanov. The authors are especially grateful to Xi Wu, who proposed to consider the system discussed in Sect. \ref{SectE2}.

\end{document}